# Possible Energy Gain for a Plasma Liner-Driven Magneto-Inertial Fusion Concept

C. E. Knapp and R. C. Kirkpatrick
Los Alamos National Laboratory, Los Alamos, NM 87545, USA

**Abstract:**

A one dimensional parameter study of a Magneto-Inertial Fusion (MIF) concept indicates that significant gain may be achievable. This concept uses a dynamically formed plasma shell with inwardly directed momentum to drive a magnetized fuel to ignition, which in turn partially burns an intermediate layer of unmagnetized fuel. The concept is referred to as Plasma Jet MIF or PJMIF. The results of an Adaptive Mesh Refinement (AMR) Eulerian code (Crestone) are compared to those of a Lagrangian code (LASNEX). These are the first published results using the Crestone and LASNEX codes on the PJMIF concept

**Introduction:**

Magneto-Inertial Fusion (MIF) targets were proposed a few decades ago as a result of experiments at Sandia National Laboratory in Albuquerque, NM [1, 2]. These "Phi targets" provided a low level of neutron yield, which was attributed to a magnetic field produced by a current along the axis of the target. An analysis that followed [3] indicated that the level of neutron yield obtained should have been expected, and a concurrent parameter study indicated that with proper design, high gain was possible [4]. Later studies of the parameter space for magnetized targets [5] indicated that the "Phi targets" operated in a yield valley, and either higher or lower fill density would have resulted in higher yield.

In the 1990's Francis Thio proposed the use of plasma jets to create a plasma liner that would drive a magnetized target plasma to fusion ignition conditions [6]. The basic idea was based on experiments at the Air Force Research Laboratory (AFRL) in Albuquerque [7], and the concept of impact fusion [8]. The goal of impact fusion was to achieve fusion by accelerating a projectile into a fusion fuel in order to heat and compress it to fusion conditions. The early proposals involved a cold hypervelocity slug impacting a conical cavity containing a fusion fuel, but it became clear that the loss of energy into the cavity walls and hydrodynamic distortion of the cavity would prevent attainment of fusion conditions. This is avoided in Thio's proposal due to the three-dimensional (3D) symmetry of the configuration.

The experiments at AFRL showed that it is possible to use multiple plasma jets to form a cylindrical plasma shell [7]. Plasma Jet Magneto-Inertial Fusion (PJMIF) [9] proposes using plasma jets to form a spherical shell with inward momentum that compresses and heats a preformed magnetized plasma target inside. In principle there is no particular requirement for the shell or target to be spherical, but the shell must be capable of providing 3D compression and confinement. Magnetization of the target dramatically reduces the electron thermal conduction, thus greatly mitigating heat loss from the fusion fuel, which in turn leads to greatly reduced power requirements for the driver. Also, if a sufficient magnetization parameter (field times radius, BR) ensues, the energy deposition by charged fusion reaction products (e.g., DT alpha particles) will be enhanced, leading to higher gain than otherwise.

Studies are currently being performed on the Plasma Liner Experiment (PLX) at LANL, which is based on the PJMIF concept, and consists of a 9ft diameter spherical vacuum chamber designed to hold up to 60 plasma guns,, and presently has two plasma guns. High efficiency of the pulsed-power-driven jets is a key advantage of this concept. The electrical-stored energy to liner kinetic energy efficiencies of 40% to 60% appear possible [10, p.7]. The efficiency of plasma guns creating and launching the jets can be optimized for a given velocity. Several recent studies have investigated various aspect of PJMIF: [11-21]. Methodical experimental efforts were initiated at Los Alamos National Lab (LANL) [22,-24] in collaboration with Hyper V Technologies [10, 25], and the University of New Mexico [26]. Basic plasma shock studies are covered in Refs. [23,24]

The present parameter study analyzes an MIF test case proposed by Thio [27] (see Table I), which is based on PJMIF. This case was identified by Thio after an extensive study using a 1D hydrodynamic code with simple models for burn and alpha particle deposition. Starting with his original configuration, a few code features were varied in a systematic way, such as resolution and front tracking, to see how the gain would be affected. The velocity of the imploding liner is the main variable of interest in this study. Other variables and features of the codes were checked out in a preliminary study, such as heat conduction and target densities, and those will be discussed briefly. After the preliminary study was complete [28], a few interesting cases of varying the velocities were run on the ICF code LASNEX [29] to compare to the Crestone AMR results, which is covered in this Letter.

**Original Configuration**:

Thio's original test case consists of several imploding 1D spherical shells. The three outer xenon shells each have the same inwardly directed velocity, but different densities (see Table I), which taken together is referred to as the liner. Thio did this to approximate a radially varying density profile in the liner. These spherical Xe shells provide the driving force. Inside the liner is a thin shell of deuterium and tritium (DT1), and inside the DT1 shell is a spherical DT target of lower density (DT2). The Xe liner and thin DT1 shell are given inward velocities of 60 km/sec and 58.8 km/sec, respectively. The DT2 target is warmer and less dense than the DT1 shell and is initially moving inwardly at 6 km/sec, but was initially static for the calculations in this study. In Thio's configuration there was a void at the origin of radius 1.0e-3 cm, which was ignored for this study. Thio assumed that the spherical liner was preformed by the merger of an unspecified number of plasma jets directed radially inward toward the origin, and the calculation starts just after an idealized merger into a spherical shell with radially inward velocity.

**Table I**
**Test Case Initial Conditions**

| Shell | Outer Radius | Density | Temperature | Velocity |
|---|---|---|---|---|
| | cm | gm/cc | eV | cm/sec |
| Air | 32.0 | 1.0e-8 | 0.025 | 0.0 |
| Xe1 | 7.697700 | 1.036878e-2 | 1.37871607 | -60.0e+5 |
| Xe2 | 6.531033 | 1.537440e-2 | 1.37871607 | -60.0e+5 |
| Xe3 | 5.364367 | 3.575443e-2 | 1.37871607 | -60.0e+5 |
| DT1 | 4.197700 | 7.763113e-4 | 0.49997846 | -58.829e+5 |
| DT2 | 4.062300 | 1.776236e-5 | 79.9655321 | -6.0e+5 |

**AMR Code:**

The calculations were performed using a computer code based on Eulerian hydrodynamics and an Adaptive Mesh Refinement (AMR) algorithm [30, 31]. All the cases considered here were run in 1D with the 3-T (3 temperatures: electron, ion and radiation with Maxwellian and Planck distributions, respectively) and radiation diffusion packages on. A fixed zoning with fine resolution was set up around the origin out to 1.5 cm, and AMR was used elsewhere, which shortened the total run time. Also, the burn package included H, D, T, He3, and He4 as fusion products. Sesame equations of state (EOSs) were used (see Table VII in the Appendix). For Eulerian calculations voids cause problems, so low density air surrounded the outer Xe shell. The Air of Table I was not part of Thio's original setup.

There are features of the code that have been turned on that also have an effect on the gain. Two features are front tracking packages, VOF and IP. These reduce the numerical diffusion of one material into another. VOF does not allow diffusion across material interfaces and was used for this study for comparison to the Lagrangian code LASNEX, which also does not allow diffusion across material interfaces. It might be more realistic to use IP, which allows some diffusion, but much less than numerical diffusion would cause. VOF increases the gain, presumably, because it does not allow any contamination of the DT by the Xe liner. VOF was applied only to the interface between the DT shell and the Xe liner.

The AMR maximum level of refinement allowed is set at the beginning of the calculation. Starting with low levels of refinement, the gain increases as the level is increased, but converges for sufficiently high levels of refinement. So, it is necessary to increase the allowed AMR refinement until convergence is achieved. A resolution study should be conducted for each new MIF target design to insure that convergence has been achieved. For this study the preliminary work was done with a minimum cell resolution of about 19.5 microns, because the cases run much faster there. However, the regime where the gain actually leveled off was around 2.4 microns minimum cell size near the origin with VOF on. (In some of the tables and the text the terms L11 and L14 are used and refer to the maximum level of resolution. L11 refers to about 19.5 micron, whereas L14 implies about 2.4 micron cell size.) However, in trying to refine to the 1.2 micron level (or L15), some of the cases do not run to completion. Some of the 1.2 micron cases would run through peak gain, and then crash but the gain did not change significantly from the 2.4 micron cases. Therefore, it appears that the calculations using 2.4 microns are reasonably converged.

A common feature of ICF codes is the ability to turn on and off (or to adjust the effect of) various physical processes by using switches or multipliers. For the AMR code electron Heat Conductivity (e-HC) can be turned on or off. Turning it on diffuses the energy from a region of higher temperature to one of lower temperature. Setting the e-HC to off, causes the electron energy to be deposited locally, and was used in Ref, 28 to approximate the presence of a strong magnetic field.

**The AMR Study:**

The main objective of this study is to find which variations of the initial input parameters increase the fusion gain, where the gain is defined as the total thermonuclear energy-out divided by the initial kinetic energy of the spherical Xe shell, which represents the merged plasma jets. The changes to Thio's case that had the most influence on gain were reducing the number of Xe layers to just one, increasing the implosion velocity of the liner, and running with e-HC on and off [28]. Reducing the number of Xe layers in the liner directly reduces the input energy more than it reduces the fusion yield, hence enhancing the gain. So far, we have not varied the thickness or density of the remaining Xe layer (designated as Xe3 in Table I), which for this study constitutes the liner.

## Table II
### Parameter Changes Made to Thio's Case of Table I

| |
|---|
| 1. Removed the outer two xenon layers of the liner. |
| 2. Increased the liner velocity (70, 80, 90, 100, 110, and 120 km/sec). |

Besides these changes it was found from preliminary work that certain other features of the code were needed. To reiterate, all cases of Tables III and IV are 1D and used 3-T with radiation diffusion. Also, except for cases #1, 2, & 3, all the other cases had the ion heat conduction (i-HC) and VOF on, and the resolution around the origin set to 2.4 microns minimum cell size in order to resolve the region of maximum compression.

## Table III

**Selected Results from the Preliminary Study for 60 km/sec**

| Case | Initial Conditions | | | TN Energy Produced | Initial Energy Kinetic | Gain |
|---|---|---|---|---|---|---|
| | | | | ergs | ergs | |
| 1. | No HC | VOF off | L11 | 2.0081e+15 | 4.9993e+14 | 4.017 |
| 2. | i-HC only | VOF off | L11 | 6.7568e+14 | 4.9993e+14 | 1.3515 |
| 3. | e-HC only | VOF off | L11 | 7.9689e+12 | 4.9993e+14 | 0.01594 |
| 4. | i-HC only | VOF on | L14 | 2.9468e+15 | 4.9993e+14 | 5.894 |

The four cases of Table III are a few of the preliminary cases with interesting parameter changes. These four are all run with the conditions specified in Table I (with the DT2 velocity set to zero), but with different physics packages turned on or off. That is, all four cases were run at 60 km/sec and with all three Xe shells. The first three start with the lower resolution of 19.5 microns cell size (L11) around the origin, and they also have VOF turned off. They are studies of turning the ion and electron heat conductions on or off. Case #1 of Table III is Thio's case with both heat conductions off. The second case has i-HC on only, and the third one has only e-HC on. The table shows that the e-HC kills the gain, and the i-HC dropped the gain some, but it was still above unity.

As a result of the first three cases, the fourth was run with i-HC on, but e-HC off, which finds the upper limit in the gain as if a strong magnetic field were imposed to eliminate electron thermal conduction. For more cases with the e-HC on, which models the case with no magnetic field, see ref. [28]. Case #4 had VOF on, and increased resolution of 2.4 micron cell size around the origin. With this, the gain increases to 5.894. This is the highest gain using the original configuration of Table I.

The 7 cases of Table IV were run with the changes outlined in Table II. Table IV shows the gains for several Xe liner velocities. By eliminating Xe1 and Xe2, and keeping only Xe3 as the liner, the gain jumps up to 12.56, because the initial kinetic energy is reduced more than the TN energy produced (see the first entry for 60 km/sec.)

The following Table IV shows the gains for different initial implosion velocities, with the two outer Xe shells removed. These cases are based on case #4 of Table III (i HC on, VOF on, and L14, which is the minimum cell size of 2.4 microns). A plot of Table IV is included below as one of the curves (squares) in Fig. 1 in the section on the comparison of the two codes.

**Table IV**

**Gains of the Parameter study with e-HC off,
i-HC on, Xe3 only, VOF on, & L14**

| Case | Initial Velocity | TN Energy Produced | Initial Energy Kinetic | Gain |
|---|---|---|---|---|
| | km/sec | ergs | ergs | |
| 1 | 60 | 2.7268e+15 | 2.1714e+14 | 12.56 |
| 2 | 70 | 6.9670e+15 | 2.9555e+14 | 23.57 |
| 3 | 80 | 1.1031e+16 | 3.8602e+14 | 28.58 |
| 4 | 90 | 1.4795e+16 | 4.8856e+14 | 30.28 |
| 5 | 100 | 1.9379e+16 | 6.0316e+14 | 32.13 |
| 6 | 110 | 2.0601e+16 | 7.2982e+14 | 28.22 |
| 7 | 120 | 2.2719e+16 | 8.6855e+14 | 26.16 |

In Tables III and IV the gain is simply the total thermonuclear (TN) yield divided by the initial kinetic energy. The highest gain was 32.13 (see Tables IV & V, and Fig. 1). This occurs for an initial velocity of 100 km/sec and e-HC off. For the AMR code the TN charged particle (CP) energy is deposited locally and then diffused, and the neutrons escape (i.e., none of their energy is deposited). The CP deposition represents an upper bound to the ion energy that a magnetic field might retain.

Note that the gain starts dropping above about 100 km/sec as the velocity increases. This is because the initial kinetic energy increases with increased velocity, but the increase in yield does not keep up. Hence 100 km/sec appears to be close to the optimal velocity for the modification of Thio's configuration (with Xe1 & Xe2 removed.)

Knapp's LA report [28] studied the effect of turning the e-HC package on and off to bracket the effects of a magnetic field. Basically, that parameter study indicated that a magnetic field is needed for liner velocities below 100 km/sec in order to reduce e-HC to an acceptable level. For 100 km/sec and greater it appeard to make little difference whether e-HC is on or off (which is unexpected and might not be realistic – deferred to the Discussion section). In any case a sufficiently strong magnetic field should allow the gains in Table IV to be reached. That study also varied the target density DT2, but that is not covered in this letter.

### LASNEX Runs:

LASNEX is a versatile ICF code [29] that has been used extensively in the past to design and analyze small fusion targets and other laser driven experiments. It is a moving grid-based Lagrangian code with magnetohydrodynamics (MHD) capabilities. Because it uses a logical mesh, the material interfaces in LASNEX are well defined. Therefore, interface algorithms like VOF and IP are not needed to suppress mix at interfaces. As noted above, in the physical setting, some material diffusion is likely to occur at interfaces (as well as within each material), but that must be the subject of another PJMIF study.

A limited number of LASNEX runs were used to confirm some of the AMR results. Cases were run with: (1) zero electron heat conduction (equivalent to e-HC off), (2) full electron heat conduction, and electron heat conduction suppressed by a B-theta field concurrently compressed along with the DT during the convergence of the Xe liner. These were done for four different inward velocities (60, 80, 100, & 120 km/sec). All the LASNEX cases were run only with the initial densities of Table I.

While LASNEX is 2D typically, lD spherical ICF calculations use an axially symmetric conical section with reflecting boundary conditions. However, for our 1D LASNEX runs, which use an azimuthal magnetic field, it was necessary to model the 1D problem as an equatorial wedge with an imposed initial magnetic field that ramped up from zero on axis to about 100 kG near the outside of the DT shell (DT1). This idealized 1D configuration provides a reasonable approximation to the more desirable 2D configuration and takes much fewer computational resources.

Gains above 30 were achieved (see Table V), but no attempt was made to alter the initial configuration in seeking higher gains. Only the initial velocities were varied. While laser driven fusion energy targets must perform with much higher gains to overcome the inefficiency of the laser driver, the efficiency associated with a plasma gun is much higher, so the acceptable target gain for PJMIF is much lower. In principle, PJMIF targets do not need gains greater than about 30 to support a viable energy system.

The LASNEX results generally agree with the Crestone AMR results for e-HC off, but for full electron heat conduction the AMR code gave much higher gains at Xe liner velocities of 100 km/sec and greater [28]. The gains for the LASNEX runs with zero e-HC and full e-HC suppressed by a B-theta magnetic field were nearly the same, suggesting that modeling these cases with zero electron heat conduction is a reasonable alternative to using an MHD code.

Magnetized targets operate far from the typical conditions found in ICF simulations. Here the magnetized fuel ignited before stagnation occurred (as is generally necessary for inertial fusion), but with areal densities far less than required for typical inertial fusion targets (a $\rho$-R ~0.03 gm/cm^2 in DT2 and DT1, for a total of ~0.06 gm/cm^2 at maximum compression). The convergence (initial over final outer radius of DT1) at maximum compression was about 17, and the density was somewhat greater than 1 gm/cc in DT2, and just a few gm/cc in the DT1 shell. The peak ion temperature exceeded 80 keV in DT2, while the radiation temperature never exceeded 1 keV. Near maximum compression, the magnetic field varied greatly within DT1 and DT2, but the $\beta$ grew from an initial value of about 4 at the outset of the calculation to 1000 in DT1 to 25000 in DT2 at the locations of the peaks in the field (from one to several MG). While the MHD includes higher level effects, such as the Nernst effect, the 1D calculations are not expected to support such effects. In a 2D calculation such effects could possibly influence the results.

One cannot take the 1D results at face value, because in the presence of a $B_\theta$ magnetic field the 2D configuration would allow enhanced energy loss along the axis, thus reducing the fusion burn inside the equatorial region. Reducing the 1D gain by a factor of about two might provide a rough estimate of the gain for a 2D calculation.

### Table V
### Summarizing the Gain Comparison Results for the Two Codes

| Velocity km/sec | LASNEX Gain | Crestone Gain |
|---|---|---|
| 60 | 0.98 | 12.56 |
| 70 |  | 23.57 |
| 80 | 32.0 | 28.58 |
| 90 |  | 30.28 |
| 100 | 33.1 | 32.13 |
| 110 |  | 28.22 |
| 120 | 28.0 | 26.16 |

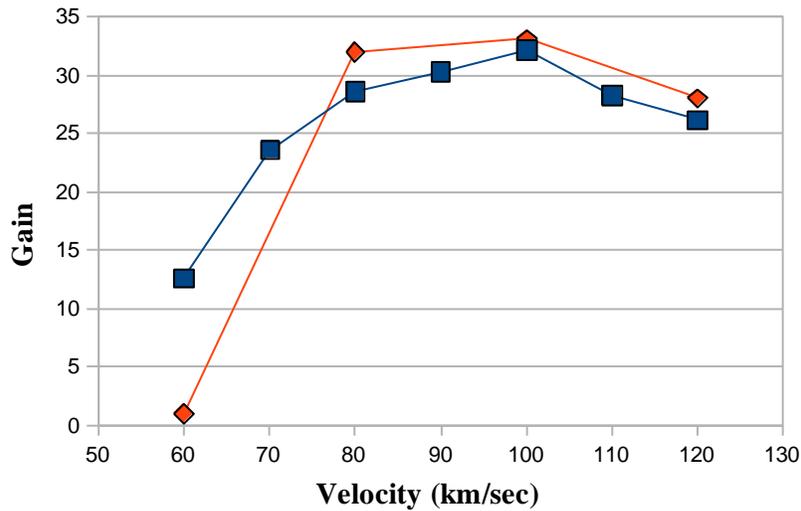

**Gain Comparison: LASNEX to Crestone**

**LASNEX – Diamonds, Crestone – Squares**

Figure 1. Plot of the gains versus the velocities of Table V comparing the results of the LASNEX calculations to those of the Crestone AMR code calculations.

**Discussion:**

While the gray diffusion was used for the radiation in the AMR code, multi-group diffusion was used in the LASNEX runs. This may be the source of some differences between the two codes. In addition, the AMR code used local energy deposition, which could lead to over prediction of fusion ignition and gain. Then the heat is diffused by thermal conduction. This was not the case for the LASNEX calculations. The LASNEX runs used a multi-group Fokker-Planck model for the transport of charged fusion reaction products, which allows non-local deposition of the fusion energy. Similar gains in Table VI suggest that the use of local deposition of charged fusion products is a reasonable approximation in the AMR cases.

In addition, for the case of 100 km/sec, LASNEX was run with both zero electron thermal conduction and with full electron thermal conduction, but with an initial azimuthal magnetic field of about 10 Tesla just inside the DT shell (DT1). Table VI shows the three 100 km/sec cases. Ein is the initial kinetic energy of the imploding Xe3 and DT1 shells, and Eout is the thermonuclear energy produced.

### Table VI
### Comparison of LASNEX and AMR Gains for 100 km/sec
### and e-HC on and off, and with a B-theta Field

| Note | LASNEX results | | | Crestone |
|---|---|---|---|---|
| | Ein | Eout | Gain | AMR Gain |
| | J | J | | |
| e-HC off | 60. | 1992. | 33.1 | 32.13 |
| e-HC on | 60. | ~6. | ~0.1 | 1.31 |
| e-HC on, B-theta | 60. | 2080. | 34.7 | - |

The PJMIF concept falls in the vast region of fusion parameter space between ICF and magnetic confinement. A major advantage of PJMIF is the dynamic formation of the fusion target, which potentially avoids the "kopec problem" (i.e., cost of target fabrication) often cited as a major obstacle for inertial fusion. Secondly, because it requires far lower implosion velocities, it allows the use of drivers with much lower power capabilities, such as pulsed power machines, which in turn provide much cheaper energy. Formation of the magnetized target inside the plasma liner must be accomplished prior to the liner formation, so that the magnetized target must have a sufficient lifetime. This is conceptually possible, but a significant experimental campaign is essential to prove the practicality of the concept.

Siemon and Lindemuth [32] have argued that the development path for magnetized target fusion (MTF, a general term that includes PJMIF) is far less expensive than for other fusion energy concepts, partly because it is not on the cutting edge of technology, and partly because it requires far less energy than MFE and far less power than ICF.

**Further work for PJMIF:**

The main parameter of this study has been the liner velocity. Other parameters of interest might be density, composition and mass of the Xe liner. The temperatures of the three gases (Xe, DT1,and DT2) should also be studied. It is possible that higher gains than those shown in Table IV may be possible for this simple configuration.

While VOF does not allow the diffusion of one material into another, in an experiment the Xe probably would diffuse into and contaminate the DT shell. This study could be extended to explore the effect of Xe contamination. Use of IP front tracking package could be used to get a sense for the effect of diffusion, since VOF stops all diffusion, and IP does allow some.

Two- and three-dimensional AMR runs are expensive for this problem, but a few have been attempted at much less resolution. The geometries for 30 and 60 merging jets have been setup without the radiation and 3-T packages, which is probably not important for the merging of the jets [33]. The code used was a Smooth Particle Hydro (SPH) code called SPHINX. The cases run, started with individual jets to study the merging, and the results displayed in Figs. 4 and 6 of Chapter V in Ref. 33 show the 2D and 3D implosions run to maximum compression. The outer edge of the liner stagnation surface is reasonably spherical, but the boundary between the liner and the target material is very distorted probably by the Rayleigh-Taylor (RT) and Kelvin-Helmholtz (KH) instabilities. However, the core of the target material is still reasonably spherical. Also there is still liner material streaming onto the stagnation surface even at maximum compression maintaining pressure on the target material. These results are good for analyzing the merger of the jets. The same geometrical setup for the SPHINX code has also been used in the Crestone codes with similar results. To do the gain calculations, however, requires much higher resolution and a lot more physics packages turned on.

Cassibry et. al. [15] have also done analysis of the RT instability, using an SPH code similar to SPHINX, and have found that "The processes of plasma liner formation and implosion on vacuum were shown to be robust against Rayleigh-Taylor instability growth." Further work is required to investigate the RT instability during deceleration of the liner against the compressed target.

Because of the higher resolution required it is best to start with the liner already formed and run in 1D spherical. It would be easy to make the appropriate modifications for the 2D or 3D geometries, since a majority of the set-up work has been done. However, because of the time involved to run at such high resolutions, the 2D and 3D calculations must be deferred until later.

An obvious problem exists for Thio's test case to be used as an actual PJMIF fusion scheme. A way needs to be found to assemble the DT target, the DT shell and the Xe liner to physically obtain the initial conditions for the current test case. One possibility might be to backfill the chamber with DT gas at the density and temperature of the target DT, then let the Xe jets merge and form a liner that sweeps up the DT ahead of it. It may be too expensive to use tritium this way, so maybe the chamber could be purged with just deuterium and then just the right amount of tritium injected ahead of the Xe liner, or in the center of the chamber as the Xe liner is sweeping up the deuterium of the chamber. This would also be a way of purging the chamber of contaminating air.

## Conclusion:

Cases with and without electron heat conduction (e-HC) have been run in an attempt to bracket similar cases with and without MHD. The magnetic field in an MHD calculation should significantly retard the escape of thermal energy. Indeed, it appears that below 80 km/sec the gain is greatly diminished for the AMR calculations with e-HC on. LASNEX agrees. For the AMR calculations, as the liner velocity increases to 100 km/sec and above, the gains for the cases with e-HC on suddenly jump to near those with e-HC off. This does not agree with the LASNEX calculations. However, for LASNEX at the lowest liner velocity (60 km/sec, with or without e-HC or a magnetic field) only low gain or no gain was achieved. Gains up to 60 were obtained in [28] with reduced density of the target DT (DT2), but these are in question and have not been verified with other codes yet.

In conclusion, the main result is that for the Xe liner velocities above 60 km/sec a magnetic field reduces the energy loss by electron thermal heat conduction (e-HC) sufficiently to allow fusion ignition to be obtained with ensuing gains that are thought to be sufficient for a PJMIF-based energy producing reactor. A further conclusion is that the two codes are in reasonable agreement for this parameter space. This is the first published result using the Crestone and LASNEX codes on the PJMIF concept, and this Letter is intended to motivate further detailed parameter studies, and spur on further investigation of this concept.

## Appendix:

Appendix A: Details for the AMR Calculations.

**Table VII**
**The Sesame Numbers Used in all the Cases of Tables III & IV**

| Region | EOS | Opacity | Conductivity |
|---|---|---|---|
| Air | 5030 | 15030 | 25001 |
| Xe3 | 5189 | 15189 | 25189 |
| DT1 | 5271 | 15251 | 25251 |
| DT2 | 5271 | 15251 | 25251 |

## Acknowledgements:

We gratefully acknowledge Dr. Francis Thio and Dr. Scott Hsu for discussions and the test case used as a starting point. We also appreciate the help from Dr. Charles Wingate and the rest of the Crestone team at the LANL. Thanks are also due Dr. Ian Tregillis (LANL) and the LLNL team for LASNEX support. This work was supported by DOE contract # DE-AC52-06NA25396.